\documentclass[3p,times]{elsarticle}

\usepackage{ecrc}

\volume{00}

\firstpage{1}

\journalname{Nuclear Physics A}

\runauth{}

\jid{nupha}

\usepackage{amssymb}

\usepackage[figuresright]{rotating}

\begin{document}

\begin{frontmatter}



\dochead{}

\title{Direct Photon Measurements at PHENIX}


\author{Baldo Sahlmueller (for the PHENIX collaboration)}


\begin{abstract}
PHENIX has measured direct photons in $p$+$p$ and Au+Au collisions at $\sqrt{s_{NN}}=200$\,GeV. The $p$+$p$ measurement is in good agreement with pQCD expectations. The nuclear modification factor in Au+Au is found to be consistent with unity for all centralities, and is compared to theoretical models. The data are found in agreement with models that take into account initial state modifications, as well as with those that account also for final state modifications.
\end{abstract}


\end{frontmatter}


\section{Introduction}
\label{sec:intro}

Direct photons are a powerful probe for studying the hot and dense matter created in ultrarelativistic heavy-ion collisions. They are produced in every stage of the collision~\cite{Turbide:2007mi,Vitev:2008vk}. In initial hard scattering processes, mostly through $q+g \rightarrow q+\gamma$ and $q+\bar{q} \rightarrow g+\gamma$, direct photons at high transverse momentum ($p_T$) are produced. High-$p_T$ photons also come from processes such as bremsstrahlung and jet fragmentation. Furthermore, theoretical models predict the production of photons by interactions of a scattered parton with the medium under the presence of hot and dense nuclear matter. Direct photon production is also affected by modifications of the initial state of the nuclei.

In $p$+$p$, direct photons are measured in order to test pQCD. They are produced in the aforementioned hard scattering processes, and as bremsstrahlung or during jet fragmentation. Their cross section can be calculated in a theoretical framework. The $p$+$p$ measurement also provides an important baseline for understanding the Au+Au data.

\section{Analysis}
\label{sec:analysis}
The data were taken with the PHENIX detector~\cite{Adcox:2003zm} in 2004 and 2006, for the measurement in Au+Au and $p$+$p$, respectively. The analyses of the two data sets are described in detail elsewhere~\cite{Afanasiev:2012dg,Adare:2012yt}. The new data have a much higher reach in $p_T$ and offer improved systematic uncertainties. The Au+Au analysis used a statistical method. Decay photons of $\pi^0$, $\eta$, and other mesons are subtracted from the spectrum of all photons. In the $p$+$p$ case, most decay photons were  first removed by the tagging method, where photons are excluded if their invariant mass with a partner photon is within the mass range of the $\pi^0$.

\section{Results}
\label{sec:results}

\begin{figure}[t]
\begin{minipage}{0.48\textwidth}
\includegraphics[width=0.9\textwidth]{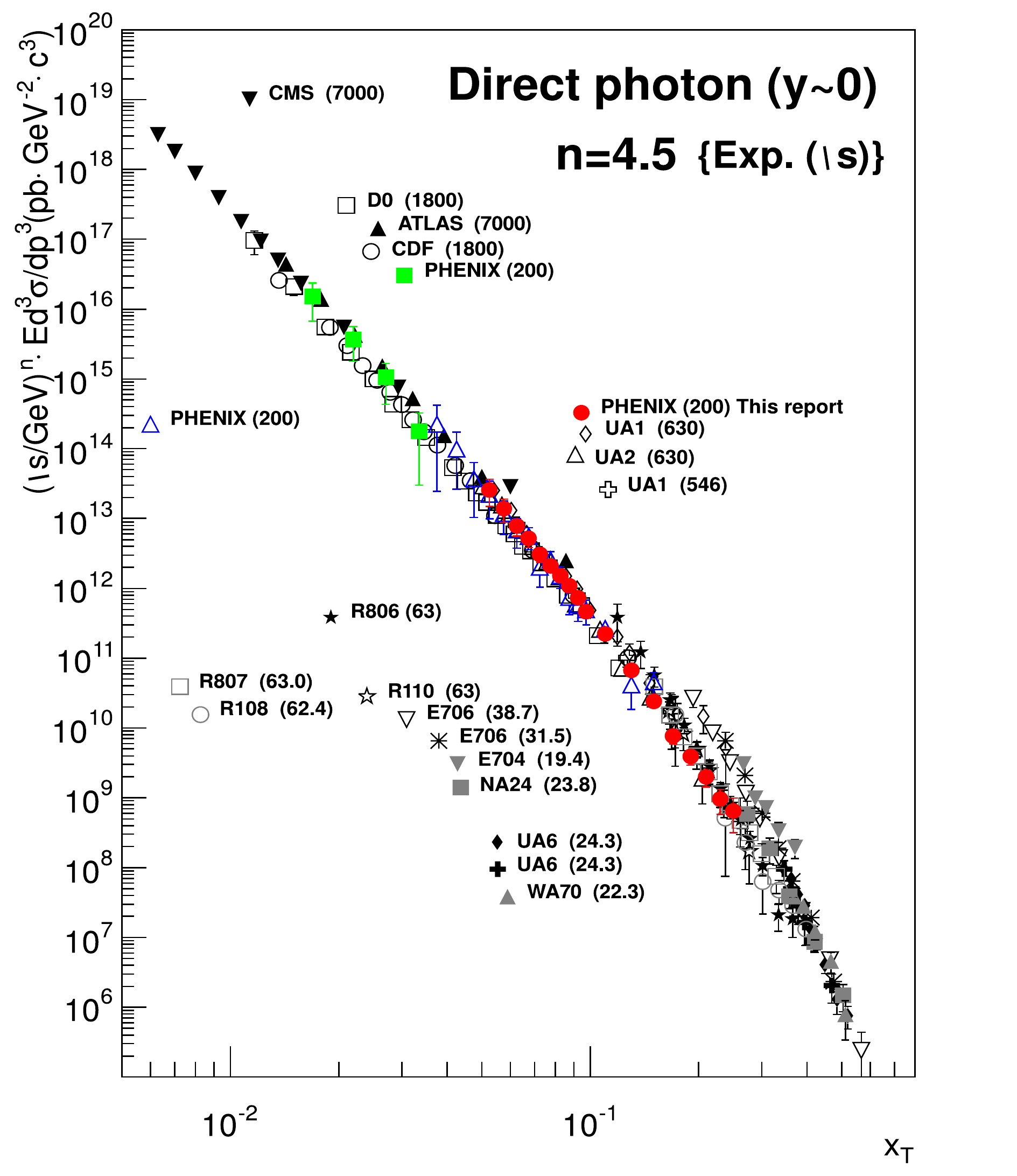}
\caption{Various direct photon cross section measurements in $p$+$p$ and
  $p$+$\bar{p}$ collisions scaled by $\sqrt{s}^{4.5}$, versus $x_T$.  The legend shows the experiment
  and the center of mass energy [GeV] in parenthesis.
\label{fig:xt}}
\end{minipage}
\hspace{\fill}
\begin{minipage}{0.48\textwidth}
\includegraphics[width=\textwidth]{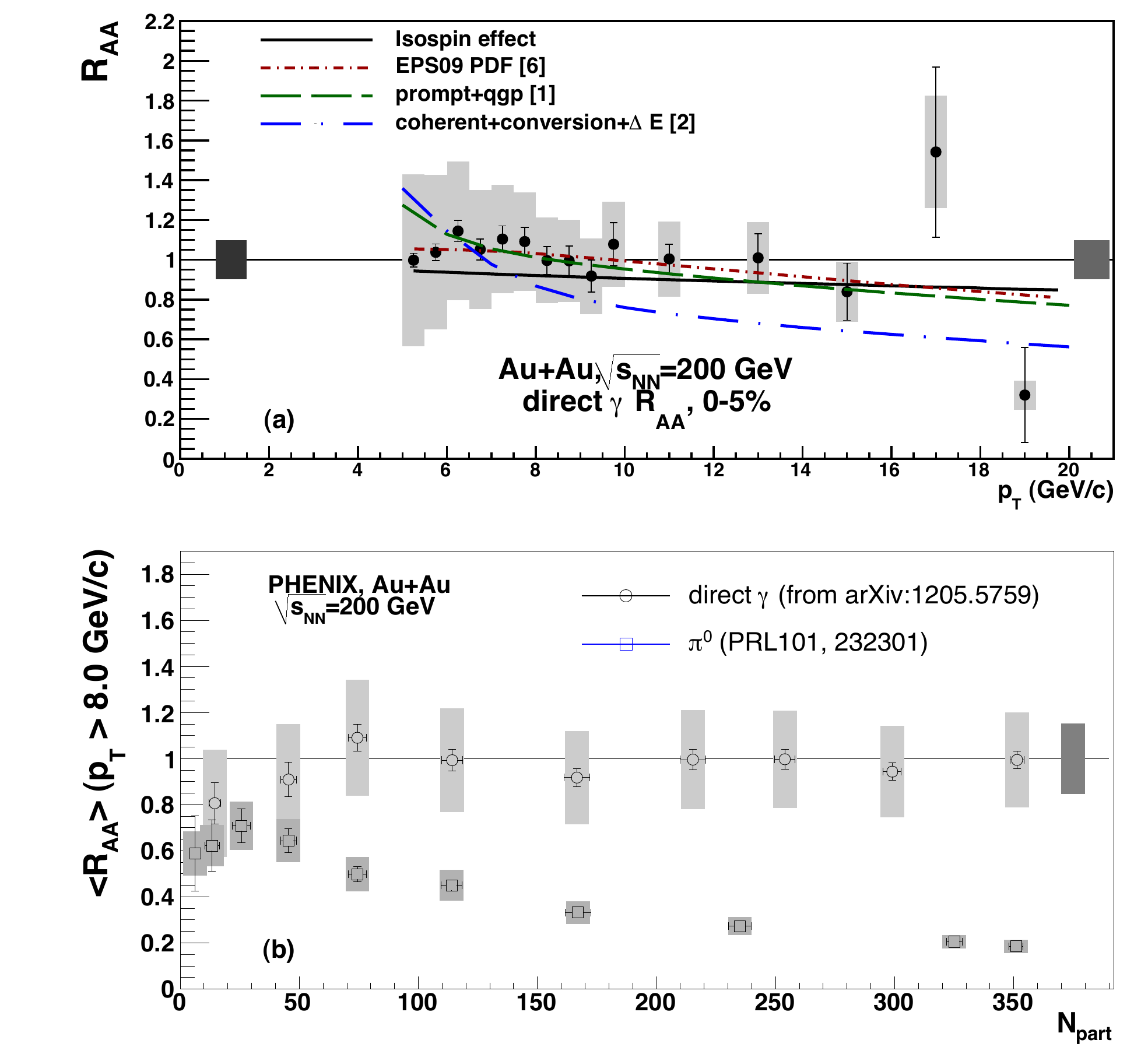}
\caption{(a) Direct photon nuclear modification factor $R_{AA}$, for 0-5\% most central Au+Au events. The data are compared with different theoretical models~\cite{Turbide:2007mi,Vitev:2008vk,Arleo:2011gc} that account for initial and final state effects. (b) Averaged nuclear modification factor $<R_{AA}>$, versus the number of participants ($N_{\rm part}$), for direct photons and $\pi^0$.
\label{fig:raa}}
\end{minipage}
\end{figure}

The high-$p_T$ direct photon cross section in $p$+$p$ collisions has been measured over a broad $p_T$ range (5\,GeV/$c < p_T <$25\,GeV/$c$)\cite{Adare:2012yt}. In Fig.\ref{fig:xt}, it is compared to many other $p+p$ and $p+\bar{p}$ measurements at various collision energies\footnote{see \cite{Adare:2012yt} for references}, as a function of $x_T = 2\cdot p_T/\sqrt{s}$. All the measurements are on one universal curve when scaling the cross section by $\sqrt{s}^n$ with a common exponent $n=4.5$. This behaviour is expected from pQCD and thus shows the validity of pQCD experimentally.

The nuclear modification factor, $R_{AA}$, was calculated as the ratio of the measured Au+Au and binary-scaled $p$+$p$ data, and is shown for 0-5\% most central events in Fig.~\ref{fig:raa}(a) and compared to theoretical models that take into account different initial and final state effects on direct photon production in Au+Au collisions~\cite{Turbide:2007mi,Vitev:2008vk,Arleo:2011gc}. The data are consistent with a scenario that accounts for a modification of the initial hard scattering cross section due to the isospin effect and modified nuclear parton distribution functions, without any final-state interactions. The existence of balancing effects of the QGP, such as suppression of fragmentation photons and enhancement from jet-medium interactions, are not excluded by the data either. However, the approach outlined in~\cite{Vitev:2008vk} is in disagreement with the data. Overall, the data show that the majority of direct photons at high $p_T$ come from hard scattering processes and suggest that possible effects from the QGP have cancelling effects.

The $p_T$ averaged nuclear modification factor, $<R_{AA}>$ is shown in Fig.~\ref{fig:raa}(b) for direct photons and $\pi^0$~\cite{Adare:2008qa}. The $\pi^0$ shows a clear suppression by a factor of ~5 towards more central events, due to jet quenching, while the direct photon is neither enhanced not suppressed.





\bibliographystyle{elsarticle-num}
\bibliography{sahlmueller}







\end{document}